# Anharmonicity of internal atomic oscillation and effective antineutrino mass evaluation from gaseous molecular tritium *β* -decay

Alexey V. Lokhov[a*], Nikita A. Titov[a]
[a] Institute for Nuclear Research RAS, Moscow, 117312
* lokhov@inr.ru

*Abstract.* Data analysis of the next generation effective antineutrino mass measurement experiment KATRIN requires reliable knowledge of systematic corrections. In particular, the width of the daughter molecular ion excitation spectrum rovibrational band should be known with a better then 1% precision. Very precise *ab initio* quantum calculations exist, and we compare them with the well known tritium molecule parameters within the framework of a phenomenological model. The rovibrational band width with accuracy of a few percent is interpreted as a result of the zero-point atomic oscillation in the harmonic potential. The Morse interatomic potential is used to investigate the impact of anharmonic atomic oscillations. The calculated corrections cannot account for the difference between the *ab initio* quantum calculations and the phenomenological model.

Keywords: molecular tritium, tritium beta-decay; neutrino mass; systematic uncertainties

PACS: 34.20.Gj; 34.50.Ez; 14.60.Pq; 23.40.Bw

## 1. Introduction

Since 1948 the tritium *β*-decay spectrum serves as a source of the most precise upper limits of the effective electron antineutrino mass $m_{\nu_e}$ [1]. A continuously improved statistical accuracy is followed by a corresponding decrease of systematic errors. Usually, systematic uncertainties are kept close to the statistical ones [2], [3].

The next-generation experiment KATRIN in Karlsruhe [4] is aimed at a sensitivity to the antineutrino mass improved by an order of magnitude. This requires a statistical uncertainty which is two orders of magnitude smaller then in the previous experiments, since the measured parameter is the antineutrino mass squared. Therefore the systematic corrections are also to be known better by at least two orders of magnitude.

The extraction of the value of $m_{\nu_e}$ from the measured spectrum of *β*-decay electrons must take into account all the losses and gains of the kinetic energy of electrons. The resulting loss of electron energy can be ascribed to the antineutrino mass provided that signature of the energy loss in beta-spectrum fits one of the non-zero antineutrino mass.

$$T_2 \rightarrow {}^3HeT^+ + e^- + \bar{\nu}_e$$
$$HT \rightarrow {}^3HeH^+ + e^- + \bar{\nu}_e \quad (1)$$

The measured decay spectrum of the molecular tritium can be derived from the bare nuclei β-decay spectrum by applying corrections that describe various effects. Some of the corrections, such as the spectrometer response function, are associated with the experimental set-up. Others are fundamental, e.g. the electrostatic interaction between the daughter nucleus and β-electron that adds the well-known Fermi-function $F(Z',E)$ to the beta-spectrum.

In this paper we consider the fundamental effect of daughter molecular ion excitation that leads to a decrease of the spectrum endpoint $E_0$. In general, the daughter molecular ion in (1) is formed in an excited state with probability described by the final states distribution (FSD). The FSD is usually presented as a set $\{\varepsilon_j, P_j\}$, where $\varepsilon_j$ is a mean value of the molecular ion ${}^3HeT^+ ({}^3HeH^+)$ excitation energy bin and $P_j$ is a total probability of transition into the given excitation energy interval. The molecular tritium decay spectrum could be presented as a weighted sum of spectra with endpoints $E_0 - \varepsilon_j$ and masses $m_{\nu_i}$ [5]:

$$\frac{dN}{dE}(E \mid m_\nu^2, E_0) = C \cdot F(Z',E) \cdot p \cdot (E + m_e c^2) \times$$
$$\times \sum_{i,j} |U_{ei}^2| P_j (E_0 - E - \varepsilon_j) \sqrt{(E_0 - E - \varepsilon_j)^2 - m_{\nu i}^2 c^4} \, \theta(E_0 - E - \varepsilon_j - m_{\nu i} c^2). \quad (2)$$

Here $E_0$ is the spectrum endpoint, $U_{ei}$ are the neutrino mixing matrix elements and $m_{\nu_i}$ are masses of the neutrino mass-eigenstates, $\theta$ is the Heaviside step function.

The main issue here is the fact that the FSD spectrum calculation has never been directly confirmed experimentally. In order to have a more model independent evaluation of the antineutrino mass in the KATRIN experiment [4] it is planned to take experimental data only within a narrow energy interval below the endpoint where only the electronic excitation ground level of the daughter molecular ion is present.

For the atomic tritium decay into the electronic ground state of the daughter ion the corresponding FSD is a single kinematics recoil line. However for the decay of the molecular tritium it becomes a sum of transitions between mother molecule and daughter molecular ion rovibrational bands, i.e. various vibrational and rotational quantum states (see [6], [7], [8]). The calculated standard deviation of the rovibrational band FSD is equal to 0.436 eV for the $T_2$ decay and 0.379 eV for the HT decay [11].

One has to point out that the mean value of the electron energy spectrum shift is not relevant. It modifies only the spectrum endpoint that is usually regarded as a free parameter of the data fit. The exact energy variation shape, however, should be known. Any unaccounted energy broadening results in a shift of the measured antineutrino mass squared that could be estimated as $\Delta m_\nu^2 \simeq -2\Delta\sigma_{FSD}^2$ [9]. Therefore, the knowledge of the FSD and a precise estimation of its width (or standard deviation) is crucial for estimating the systematic uncertainties for the $m_\nu^2$ measurement.

It could be useful to relate the FSD parameters phenomenologically with some measured properties of involved molecules. A very attractive attempt to evaluate the width of FSD of the molecular ions produced in the tritium nucleus decay HT $\to$ $^3$HeH$^+$ and T$_2$ $\to$ $^3$HeT$^+$ was presented in [10]. The Doppler-like broadening of the emitted electron's energy was calculated within a very basic model of harmonic oscillations of tritium nuclei bounded in HT or T$_2$ molecules. Due to the quantum nature of the effect the nuclei possess finite zero-point oscillations at any temperature. The Doppler broadening caused by zero-point oscillations explained most of the rovibrational band width. The remaining 4-8% discrepancy between the *ab initio* quantum calculations [6, 8] and the phenomenological model of zero-point motion was assigned to the rotational degrees of freedom of the initial molecules.

The simple rovibrational picture was discussed in [11] and a closer agreement between the *ab initio* theoretical calculations and the phenomenological description was achieved. The accuracy assumed by KATRIN for the *ab initio* calculation is 1%, and it is of interest to compare the quantum-mechanical theory with phenomenological models that are independent of structural details, with this level of precision as a benchmark. Furthermore, the quantum-mechanical calculations [6] were performed via so-called geminal-basis method and the calculations with the use of the configuration-interaction method are now in progress. However, before the confirmation of one method by another there is still room for phenomenological studies that bring further understanding of the FSD and the tritium spectrum.

In this paper we consider a more elaborate model of the (initial) T$_2$ tritium molecule that aims to narrow the gap between the phenomenological and the *ab initio* quantum calculations [6].

In Sections 2-4, instead of the harmonic potential, we use a more general Morse one [12] for a two-atomic molecule with close masses of the composing atoms. This model yields an exact solution: the rotational and vibrational spectrum and the wave functions. The parameters of Morse potential were selected to match the measured spectrum of the tritium molecule [13]. We calculate the rovibrational FSD for the Morse potential and the corresponding Doppler-like broadening of the emitted electron spectra.

For the second refinement one notices that the wave functions of the Morse potential are no longer Gaussian. This can change the momentum distribution of the initial tritium atoms and thus the shape of the β-spectrum. The impact of the non-Gaussian form of the Morse wave functions is estimated in Section 5. We discuss the results and make a conclusion in Section 6.

## 2. Doppler-like broadening for T$_2$ molecule

The Doppler-like recoil energy broadening has a purely kinematical origin. Consider the decay of a tritium nucleus in the T$_2$ molecule. The decaying nucleus has momentum $p_0$, energy $E_0$ (from this point on the subscript 0 is used for the decaying nucleus). The total energy release in the decay is $E_{max}$, the emitted electron and antineutrino have energies $E_e$, $E_\nu$ and momenta $p_e$ and $p_\nu$ correspondingly. The daughter nucleus has energy $E_1$ and momentum $p_1$.

The momentum conservation defines the nuclear recoil momentum $\vec{q}$

$$\vec{p}_0 - \vec{p}_1 = \vec{q} = \vec{p}_e + \vec{p}_\nu \tag{3}$$

Using energy conservation one obtains for the energy of the emitted electron and antineutrino:

$$E_0 + E_{max} = E_1 + E_\nu + E_e$$
$$E_e + E_\nu = E_{max} - E_R \tag{4}$$

where $E_R$ is a recoil energy transferred to the molecular ion $^3$HeT. Though the relativistic corrections to the tritium beta-spectrum considered in [6, 8, 11] can be of the order of magnitude of 1%, they come from the movement of the emitted electron, while the nuclei motion remains purely non-relativistic (with accuracy about $10^{-5}$). Therefore, one can refrain here from the relativistic energy-momentum relations for the recoiled nuclei. The non-relativistic movement of the nuclei allows one to rewrite $E_R$ using (3) and $E_0 = \dfrac{p_0^2}{2m_a}$, $E_1 = \dfrac{p_1^2}{2m_a}$, in the form of

$$E_R = \frac{(\vec{p}_0 + \vec{q})^2}{2m_a} - \frac{\vec{p}_0^2}{2m_a} = \frac{\vec{q}^2}{2m_a} + \frac{\vec{p}_0 \vec{q}}{m_a} \tag{5}$$

Since the initial nuclei momentum $p_0$ in the gas molecules T$_2$ is not constant but is distributed with some probability density $n(\vec{p}_0)$ the recoil energy is not fixed and has distribution $S(E_R)$. This distribution implies that transferred to the leptons energy ($E_{max} - E_R$) is not constant. In scattering experiments this function usually is called a scattering function. It can be written in the following form:

$$S(E_R) = \int n(\vec{p}_0) \delta\left( E_R - \frac{\vec{q}^2}{2m_a} - \frac{\vec{q}\vec{p}_0}{m_a} \right) d\vec{p}_0 \tag{6}$$

Note that at the endpoint of the decay electron spectrum, the absolute value of the recoil momentum $q$ is determined completely by the reaction energy release $E_{max}$. The influence of antineutrino escape near the spectrum endpoint has been estimated in [11] and appears to be small.

A phenomenological description of the electronic ground level excitation part of the FSD could be obtained by choosing the momentum distribution of decaying nuclei. The Gaussian form of $n(\vec{p}_0)$ that corresponds to the harmonic molecular potential yields the Gaussian form of $S(E_R)$:

$$S(E_R) = \frac{1}{\sqrt{2\pi}\sigma} \exp\left\{-\frac{1}{2}\frac{(E_R - \bar{E}_R)^2}{\sigma^2}\right\} \tag{7}$$

where $\bar{E}_R = \frac{\vec{q}^2}{2m_a}$ and

$$\sigma = \sqrt{\frac{4}{3}\langle E_{kin}\rangle \bar{E}_R} \tag{8}$$

With the known mean kinetic energy of decaying nuclei $\langle E_{kin}\rangle$ one can estimate the standard deviation of the excitation energy of the recoil ion $\sigma$ and compare it with the results of the *ab initio* calculation [6].

The recoil kinetic energy, as discussed in [11], equals to 3.41 eV (for the electron energy close to the tritium spectrum endpoint) since the daughter ions (HeT and HeH) act more like weakly bounded systems.

Validity of the above picture could be proven in the processes of elastic high-momentum transfer electron scattering from hydrogen (or tritium) molecules [14]. In this experiment a clear Gaussian shape of the elastic scattering peak was observed with a width corresponding to the zero-level mode of the $H_2$ molecule oscillation. One has to admit that such a picture is fully applicable only in a `quasi free regime' when the energy transfer $E_R$ is much larger than the molecular binding energy $E_B$, $E_R >> E_B$. However, in the tritium decay of $T_2$ and TH both energies are very close $E_R \approx E_B$.

### 3. Rovibrational spectrum of T2 molecule

In this section we calculate the mean value of kinetic energy of decaying nuclei $\langle E_{kin}\rangle$ based on the Morse potential [12] formalism and the experimental data [13]. The Morse potential description of the $T_2$ molecule provides a good opportunity to go beyond zero-vibrational harmonic potential estimates. It can be presented in a simple analytical form:

$$V(r) = D\left(e^{-2\alpha x} - 2e^{-\alpha x}\right), \quad x = \frac{r - r_0}{r_0} \tag{9}$$

Here $D$, $\alpha$ and $r_0$ stand for the depth of the potential, its sharpness and the equilibrium distance between atoms, correspondingly.

The Morse potential transforms into the harmonic potential in the limit of small deviations from the equilibrium, thus, for the lower energy levels.

So, if $|x| \ll 1$, than $V(r) = D(-1 + \alpha^2 x^2 + ...) \approx -D + \frac{1}{2}\mu\omega^2 (r - r_0)^2$, were $\omega^2 = \frac{2D\alpha^2}{\mu r_0^2}$ and $\mu$ stands for the reduced mass of the molecule. This is an excellent reference point which allows one to compare results with zero-point estimates [10, 11].

The excitation energy spectrum of the molecule is a starting point to estimate the mean kinetic energy $\langle E_{kin} \rangle$ of the decaying tritium nucleus. The Schrödinger equation has an analytical solution for the Morse potential:

$$\chi_0(x) = \frac{1}{N} y^{\beta/\alpha} e^{-\frac{1}{2}y} {}_1F_1(a, c; y) \tag{10}$$

Here ${}_1F_1(a, c; y)$ is a confluent hypergeometric function of the first kind, $y = \xi e^{-\alpha x}$, $\xi = \frac{2\gamma}{\alpha}$, $\gamma^2 = 2D\mu r_0^2$, $\beta^2 = -2E\mu r_0^2$, $c = 2\frac{\beta}{\alpha} + 1$, $a = \frac{1}{2}c - \frac{\gamma}{\alpha}$. Using these notations one can write the energy spectrum in a compact form:

$$E(\nu) = -D + \omega\left\{\left(\nu + \frac{1}{2}\right) - \frac{1}{\xi}\left(\nu + \frac{1}{2}\right)^2\right\} \tag{11}$$

The first term inside the braces is the usual harmonic oscillator spectrum and the second one corresponds to the anharmonic corrections.

The peculiarity of the Morse potential is that it provides a simple way of taking into account the rotational motion of atoms in the T2 molecules. For that purpose one adds a centrifugal potential

$$V'(r) = \frac{l(l+1)}{2\mu r^2} = \frac{l(l+1)}{\gamma^2} D \left(\frac{r_0}{r}\right)^2 \approx \frac{l(l+1)}{\gamma^2} D \left(C_0 + C_1 e^{-\alpha x} + C_2 e^{-2\alpha x}\right) \tag{12}$$

and obtains the wave functions and the energy spectrum via the similar procedure as that for the initial Morse potential:

$$\chi_l = y^{\beta_1/\alpha} e^{-\frac{1}{2}y} {}_1F_1(\tilde{a}, \tilde{c}; y) \tag{13}$$

$$E = \frac{1}{2\mu r_0^2}\left\{-\gamma^2 + 2\alpha\gamma\left(\nu + \frac{1}{2}\right) - \alpha^2\left(\nu + \frac{1}{2}\right)^2 + l(l+1) - \frac{3(\alpha-1)}{\alpha\gamma}\left(\nu + \frac{1}{2}\right)l(l+1) - \frac{9(\alpha-1)^2}{4\alpha^4 \gamma^2} l^2(l+1)^2\right\} \tag{14}$$

The values of the $\tilde{a}, \tilde{c}$ in (13) are modified according to the additional terms in the potential (12).

Equation (14) is a special case of the so-called Dunham series expansion of the energy spectrum of a molecule. For each electronic state of a molecule its energy spectrum can be presented as a sum $E_{\nu l} = \sum_{ik} Y_{ik} \left(\nu + \frac{1}{2}\right)^i l^k (l+1)^k$. The last term in formulae (14) can be omitted as its coefficient is two orders of magnitude smaller and the rotational states with l>2 will not be considered. The energy of molecule will be estimated by taking into account harmonic oscillations, anharmonic correction, simple rotational term and the first rovibrational term. The spectrum of the molecule then is

$$E_{v,l} = \omega_e\left(v+\frac{1}{2}\right) - \omega_e x_e\left(v+\frac{1}{2}\right)^2 + B_e l(l+1) - \alpha_e\left(v+\frac{1}{2}\right)l(l+1) + Y_{00} \quad (15)$$

where $\omega_e, \omega_e x_e, B_e, \alpha_e$ and $Y_{00}$ are conventional notations for the expansion coefficients. One way to estimate coefficients from Eq.(14) is to recalculate them from parameters $r_0$, $D$ and $\alpha$ extracted from measurements and calculations of H$_2$ molecular spectrum (see, for instance, [15]). Another way is to use the measurements and calculations of the T$_2$ spectrum itself [13, 16]. Since measurements (via Raman spectroscopy) and calculations of T$_2$ provide more reliable estimates of expansion coefficients, we will use them to get information about the rovibrational levels of T$_2$ and about the parameters of our modelling Morse potential (we use natural units). From [13, 16] one obtains

$$\omega_e = 2546.4 cm^{-1}, \omega_e x_e = 41.23 cm^{-1},$$
$$B_e = 20.335 cm^{-1}, \alpha_e = 0.5887 cm^{-1}, Y_{00} = 2.8 cm^{-1} \quad (16)$$

One can now calculate the rovibrational T$_2$ molecule level energies and estimate the broadening (7). The parameters of Morse potential were estimated from the data compilation [13]. Note that the comparison of formulae (15) and (14) provides the following relations:

$$\omega_e = \frac{\alpha\gamma}{\mu r_0^2}; \omega_e x_e = \frac{\alpha^2}{2\mu r_0^2}; B_e = \frac{1}{2\mu r_0^2} \quad (17)$$

From (16) and (17) one numerically obtains

$$\frac{1}{\mu r_0^2} = 40.67 cm^{-1}; \alpha^2 = 2.03;$$
$$\gamma = 43.97; D = 39314 cm^{-1} \quad (18)$$

These values will be used below for the evaluation of the momentum distribution of decaying T nuclei.

### 4. Estimating the Doppler-like broadening

In the recent paper [11] the standard deviation of rovibrational band (8) was estimated for the harmonic oscillator internuclear potential with account of rotational degrees of freedom (results in the first three terms in formulae (15)) . In the case of harmonic potential the virial theorem connects the lowest level of T$_2$ of molecule energy $E_{zp} = E_{v=0,l=0}$ and the mean kinetic energy of internal movement of the molecule $\frac{\langle p_0^2 \rangle}{2\mu}$. Namely, $\frac{\langle p_0^2 \rangle}{2\mu} = \frac{1}{2}E_{zp};$

$$\langle E_{kin} \rangle = \frac{1}{2}\frac{\mu}{m_a}E_{zp}.$$

In this Section the energy spectrum (14) is calculated with the parameters (16)-(18) using all the terms of formulae (15). Since the difference between Morse and harmonic potentials is a second order effect, in this Section it is assumed that the equality of the mean internal kinetic and mean potential energy stands also for Morse potential lowest energy level. This assumption is (indirectly) tested in Section 5 below when the wave functions for the Morse potential are compared with that of the harmonic potential.

We consider the first three rotational levels ($l = 0,1,2$) and only the ground vibrational level $v = 0$ since (see, for instance [11]) other energy levels are not occupied at 30K in the gaseous tritium source of the KATRIN experiment.

Table I presents the standard deviation of the Doppler-broadened FSD peak. Calculations were made both for the fraction of ortho- state of T$_2$ molecules at temperature $T = 30K$ and room temperature, the last one with the fraction of ortho- state of T$_2$ molecules $\lambda = 0.75$. Due to a frequent tritium circulation through room temperature equipment it is expected that ortho fraction will be closer to $\lambda = 0.75$ [11]. Table II shows the resulting total widths of FSD, i.e. weighted sums of standard deviations from Table I obtained via the model described in this paper. For the comparison presented are the results of the harmonic potential model [11] and the *ab initio* calculations [6].

From Table II one infers that the model with the Morse potential and the corresponding molecular spectrum that takes account of the first excited rotational terms and the rovibrational term in (15) yields smaller FSD standard deviations (the broadening widths are smaller) than those obtained in [11]. The aggregate value $\sigma_{tot}^{T_2} = 0.426 eV$ lies therefore farther from the theoretically calculated one $\sigma_{tot}^{T_2} = 0.436 eV$ [6]. It is the result of using the more precise parameters of the potential and molecular spectrum (16).

***Table I.*** *Rotational state energies from rovibrational T2 molecule spectrum (15), standard deviations of FSD peak and the population of each rotational level* [11]*.*

| Rotational level, $l$ | $E_{v=0,l}, eV$ | $\sigma_l^{T_2}, eV$ | $\sigma_l^{T_2}, eV$ [11] | $P(\%), T = 30K$, *Thermal* | $P(\%), T = 30K$, $\lambda = 0.75$ |
|---|---|---|---|---|---|
| 0 | 0.1569 | 0.422 | 0.420 | 43.7 | 24.6 |
| 1 | 0.1616 | 0.428 | 0.433 | 55.7 | 75.0 |
| 2 | 0.1718 | 0.442 | 0.458 | 0.62 | 0.35 |

**Table II.** Total FSD standard deviation for the thermal equilibrium and for $\lambda = 0.75$ ortho-para ratio.

| Total FSD standard deviation | $T = 30K, Thermal$ | $T = 30K, \lambda = 0.75$ |
|---|---|---|
| this paper | $\sigma_{tot}^{T_2} = 0.425 eV$ | $\sigma_{tot}^{T_2} = 0.426 eV$ |
| [11] | $\sigma_{tot}^{T_2} = 0.428 eV$ | $\sigma_{tot}^{T_2} = 0.430 eV$ |
| [6] | | $\sigma_{tot}^{T_2} = 0.436 eV$ |

## 5. Morse potential wave functions and momentum distribution

In this Section, the influence of anharmonicity on the standard deviation of rovibrational band is estimated.

The phenomenological estimates in the previous Section are based on the assumptions that the potential (9), (12) near the equilibrium distance resembles the harmonic potential $V(x) \sim x^2$ and the considered lowest energy level ($v=0$) has the same wave function as that of the ground state of a harmonic oscillator. The wave function of the lowest level of harmonic oscillator has a form of the Gaussian function. The probability density in coordinate and momentum representations are then also Gaussian distributions. That allows one to use the standard deviation of momentum distribution in (6) to obtain the FSD peak width (8) via the virial theorem (see Section 4).

If the momentum distribution of decaying nuclei is no longer Gaussian the FSD peak has also a non-Gaussian form and its width has to be calculated more accurately. The corresponding correction can lead to a wider momentum distribution and to a broader FSD peak (since the Gaussian distribution has the smallest width possible). One can therefore indirectly test the validity of exploiting the harmonic potential and the virial theorem in Section 4. For that one compares the momentum distributions obtained from the wave functions of the harmonic and Morse potentials.

1) Compare the wave function of the harmonic oscillator ground state $f_0(x)$ (the normalization constants are omitted):

$$f_0(x) \sim \exp\left\{-\frac{\omega m r_0^2}{2}x^2\right\} = \exp\left\{-\frac{\gamma\alpha}{2}x^2\right\}; \qquad (19)$$

with that of the ground state of the Morse potential $\chi_0(x)$:

$$\chi_0(x) \sim y^{\beta/\alpha} e^{-\frac{1}{2}y}; \quad y = \xi e^{-\alpha x}; \xi = \frac{2\gamma}{\alpha}; \gamma^2 = 2D\mu r_0^2; \beta^2 = -2E\mu r_0^2 \qquad (20)$$

For the Morse potential, the ground state of the nuclei motion is restricted to a narrow region around the equilibrium distance $r_0, x=0$. One can consider the expansion of the wave function $\chi_0(x)$ with respect to a small parameter x (we again omit the normalization constant):

$$\chi_0(x) \sim y^{\beta/\alpha} e^{-\frac{1}{2}y} \sim e^{-\alpha x \beta/\alpha} e^{-\frac{1}{2}\xi e^{-\alpha x}} \approx e^{-\alpha x \beta/\alpha} e^{-\frac{1}{2}\xi\left(1-\alpha x + \frac{\alpha^2 x^2}{2}\right)} \sim$$

$$\sim \exp\left\{-\left(\frac{\gamma\alpha}{2}x^2 + (\beta-\gamma)x\right)\right\} \sim \exp\left\{-\frac{\gamma\alpha}{2}\left(x + \frac{(\beta-\gamma)}{\gamma\alpha}\right)^2\right\} \qquad (21)$$

The approximate wave function (21) appears to have the same Gaussian form with the same width as the harmonic-potential wave function $f_0(x)$ in (19). The only difference is that wave function is no longer centered at the equilibrium distance $x=0$, its center is slightly moved to the point $x' = -\frac{(\beta-\gamma)}{\gamma\alpha} \ll 1$. For our purposes, however, the main characteristic is the standard deviation of the probability density function of the momentum distribution $n(p) = |\chi_0(p)|^2$, and it is the same as for the harmonic oscillator.

This justifies our assumption made in Section 4 that the total energy is divided equally between the mean kinetic and mean potential energy for the lowest energy states of the Morse potential as well as for the harmonic potential. Under this assumption the transition from (6) to (7) is also correct.

2) To take into account the next order (non-Gaussian) terms of the expansion of the wave function $\chi_0(x)$ one refers to numerical calculation using the estimates of the parameters of the potential (18). First, the wave functions in the momentum representation are derived via the Fourier transform.

$$f_0(p) \sim \int_{-\infty}^{\infty} f_0(x) e^{-ipx} dx \qquad (22)$$

$$\chi_0(p) \sim \int_{-\infty}^{\infty} \chi_0(x) e^{-ipx} dx \qquad (23)$$

The function $\chi_0(p)$ is no longer Gaussian and has a more complex form that includes an exponential multiplier and contains also the Euler gamma function Γ(z). The properly normalized functions $f_0(p)$ and $\chi_0(p)$ yield the probability density functions for the distribution of momentum of the nuclei in T$_2$ molecules: $n_{harmonic}(p) = |f_0(p)|^2$ and $n_{Morse}(p) = |\chi_0(p)|^2$ correspondingly. Estimating numerically the variations of the two distributions one obtains that the variance $\langle p^2 \rangle$ of $n_{Morse}(p)$ is slightly (<1%) smaller than that of $n_{harmonic}(p)$. This can be explained by a smaller total energy of the ground state of the Morse potential (due to the anharmonic corrections to the energy levels) and, thus, a smaller mean kinetic energy (see Section 4). In Fig.1 the distributions $n_{harmonic}(p)$ (violet, solid) and $n_{Morse}(p)$ (blue, dashed) are almost indistinguishable and their difference (brown, dotted) is smaller than 1% near the peak. Since the distribution $n_{Morse}(p)$ differs from the Gaussian one it is reasonable to check also possible contribution of the tails of the distribution. The difference of the calculated mean momentum square on the integration interval: $\langle (pr_0)^2 \rangle = \int_{-(pr_0)_{max}}^{(pr_0)_{max}} (pr_0)^2 n(pr_0) d(pr_0)$ is shown in Fig.2. One can infer that the tails of distributions $n_{Morse}(p)$ as well as for $n_{harmonic}(p)$ do not contribute to the overall value of mean momentum square.

The variation of mean kinetic energy $\langle E_{kin} \rangle = \left\langle \dfrac{\vec{p}_0^2}{2m_a} \right\rangle$ and corresponding rovibrational FSD peak width (8) appears to be about -1.5% and -0.8%, correspondingly.

Therefore, the impact of anharmonicity on the FSD peak appears to be negligible.

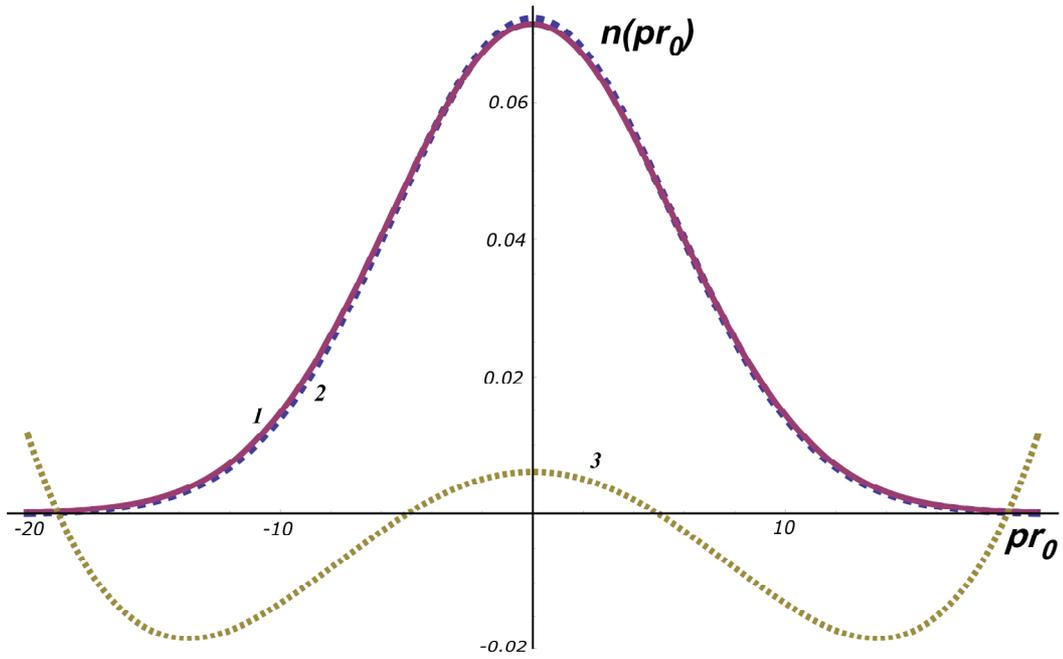

**Fig. 1** Momentum probability distributions for the harmonic potential $n_{harmonic}(p)$ (violet, solid, 1) and the Morse potential $n_{Morse}(p)$ (blue, dashed, 2). The parameters of the potentials were chosen according to (18). The normalized difference $\frac{(n_{Morse}(p) - n_{harmonic}(p))}{(n_{Morse}(p) + n_{harmonic}(p))}$ (brown, dotted, 3) is presented for the sake of comparison, $p \cdot r_0$ is a dimensionless momentum.

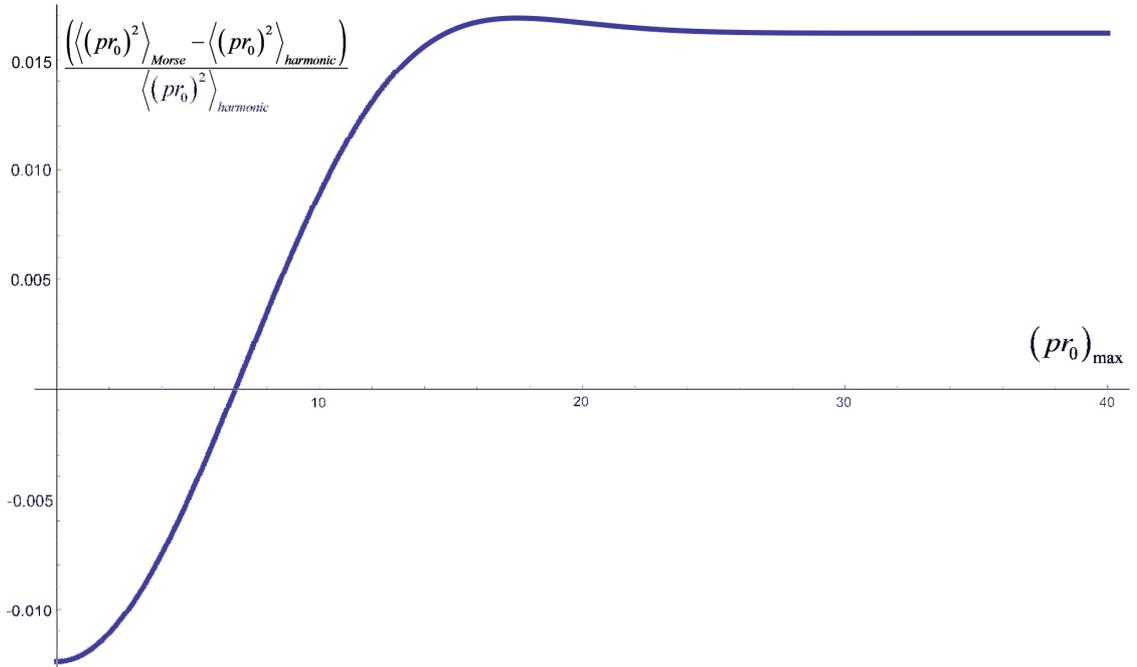

**Fig. 2** Difference of the calculated variances of the momentum distributions $n_{harmonic}(pr_0)$ and $n_{Morse}(pr_0)$ for various intervals of integration $(pr_0)_{max}$.

## 6. Discussion and conclusions

We considered a descriptive model of tritium molecule based on the Morse potential in attempt to close a gap between the phenomenological model [11] and the *ab initio* calculations [6] of the standard deviation of the final state energy distribution of the β-decay products. The Morse potential provides a transparent phenomenological picture of nuclei motion within the molecule. The corresponding rovibrational spectrum and wave functions were used to estimate the Doppler-like broadening of the final state distribution due to this motion.

The rovibrational spectrum of $T_2$ molecule was improved by matching the Morse potential parameters to the Raman spectroscopy data [16].

Provided that the Morse wave functions and potential do not differ much from the harmonic one, the more precisely calculated energy of the lowest molecular level was introduced into the formula for FSD variance (see Table 1). The calculated aggregate variance $\sigma_{tot}^{T_2} = 0.426 eV$ appears to be smaller than that obtained in the phenomenological model [11] $\sigma_{tot}^{T_2} = 0.430 eV$ and in the full theoretical calculations by Saenz et al., $\sigma_{tot}^{T_2} = 0.436 eV$ [6].

The comparison of the Morse potential wave functions and that of the harmonic potential shows that the non-Gaussian form of the Morse potential wave function does not affect much the nuclei momentum distributions $n(p)$ (see Fig.1) and thus the variance of the FSD. For the estimated parameters of Morse potential (18) and for the lowest rovibrational level the variance was calculated numerically and differs from the variance for the harmonic potential by less than 1%.

The phenomenological model of zero-point motion described by the harmonic or Morse potential provides a qualitative explanation for the width of the peak in FSD in the molecular tritium beta-decay. However, neither the harmonic potential, nor the more sophisticated Morse model of the tritium molecule yields the FSD variance close enough to the *ab initio* calculated value $\sigma_{tot}^{T_2} = 0.436 eV$. The phenomenological value falls by about 1.5% and 2.5% below the theoretical one. The phenomenological descriptions of the FSD rovibrational peak are not in a quantitative agreement with the calculations [6]. However, the most of the width of the peak is thought to be well understood and it comes from the zero-point and rotational motion of the mother tritium molecule. Notice that the main contribution to the broadening of the peak comes from the lowest part of the molecular potential (near equilibrium) where the momentum of the nuclei is the highest. Since this part of the potential can be usually described by a parabola, the harmonic potential provides the leading term and the form of the potential far from equilibrium introduces only minor corrections to the FSD.

A further complication of the model of decaying nuclei motion within $T_2$ or TH molecules seems to be unjustified since it eventually brings one to the *ab initio* theoretical calculations [6], while the aim of the studies [11] and the one in this paper was to build a reliable phenomenological picture. Besides, the discrepancy between the phenomenological results and the *ab inito* calculations [6] may naturally be connected with the limited applicability of phenomenological model in the regime when transferred energy is close to the molecular binding one [14]. In such a regime an account of the final states spectrum of the daughter molecular ions is obviously required [17]. The exploited formalism can be further applied to similar studies of systematic uncertainties of the KATRIN tritium spectrum, in particular to the scattering of electrons in the tritium source, which is a work in progress. Further studies of the final states distribution could be conducted via dedicated experiments such as TRIMS [18].

**Acknowledgements**

The authors thank Alejandro Saenz and Hamish Robertson for fruitful discussions, Fyodor V. Tkachov for careful reading of the text and interest in this work, Professor Ernst Otten for reviewing the paper and giving valuable comments and the collaborators of the KATRIN and Troitsk-$\nu$-mass experiments for discussions.